\documentclass{article}

\usepackage[margin=1in]{geometry}
\usepackage{amsmath,amssymb,amsthm}
\usepackage{natbib}
\usepackage{graphicx}
\usepackage{soul}
\usepackage{color}

\begin{document}
\title{Bayesian ``Deep'' Process Convolutions:\\An Application in Cosmology}
\author{Kelly R. Moran, Richard Payne, Earl Lawrence, David Higdon, Stephen A. Walsh, \\
Annie S. Booth, Juliana Kwan, Amber Day, Salman Habib, Katrin Heitmann}
\date{}
\maketitle

\newcommand{\Ps}{\mathcal{P}}

\newtheorem{thm}{Theorem}[section]
\newtheorem{cor}[thm]{Corollary}
\newtheorem{lem}[thm]{Lemma}

\begin{abstract}
The nonlinear matter power spectrum in cosmology describes how matter density fluctuations vary with scale in the universe, providing critical insights into large-scale structure formation.
The matter power spectrum includes both smooth regions and highly oscillatory features.
Cosmologists rely on noisy, multi-resolution realizations of large N-body simulations to study these phenomena, which require appropriate smoothing techniques to learn about underlying structures.
We introduce a Bayesian Deep Process Convolution (DPC) model that flexibly adapts its smoothness parameter across the input space, enabling it to capture both smooth and variable structure within a single framework. 
The DPC model leverages common patterns across related functions to improve estimation in regions with sparse data. 
Compared to existing methods, the DPC model offers superior accuracy and uncertainty quantification in simulated data, and qualitatively superior performance with the cosmological data.
This methodology will be useful in cosmology and other fields requiring flexible modeling of smooth nonstationary surfaces.
\end{abstract}

\section{Introduction}
The field of cosmology is concerned with studying the origin and evolution of the universe on large scales.
Small early density fluctuations near the origin of the universe lead to later formation of galaxies, galaxy clusters, and larger structures such as galaxy superclusters and filaments.
Linear structure formation theory is effective for understanding the early, less dense universe, while nonlinear structure formation theory becomes essential in the later universe as structures grow and gravitational interactions become more pronounced.
Typically, quickly-computed theoretical results can describe how cosmological parameters govern linear structure formation, while computationally expensive $N$-body simulations are needed to explore this relationship for nonlinear structure formation.

The ``Coyote Universe" and ``Mira-Titan'' projects (named after the clusters the data were simulated on, both of which have since been retired) were two simulation suites built to better understand structure formation \citep{heitmann2010coyote, heitmann2009coyote, lawrence2010coyote, heitmann2016mira, lawrence2017mira, moran2023mira}. 
Due to the computational demand of these simulations, they could only be run at a limited number of cosmological parameter settings (i.e., cosmologies).
The Coyote papers first described the construction of a statistical emulator for the nonlinear matter power spectrum.
This emulator provided an accurate and computationally inexpensive approximation to the simulation realizations of the nonlinear power spectrum for any cosmology within a range of input values.
The Mira-Titan suite aimed to broaden the coverage of the Coyote simulation campaign, adding massive neutrinos and a dynamical dark energy equation of state, expanding the considered cosmological parameter space, and substantially enhancing simulation quality in terms of both volume and resolution.
For both the Coyote and Mira-Titan suites, the smooth matter power spectrum underlying each set of noisy simulation realizations was learned using a Deep Process Convolution (DPC) model, and these smooth spectra were used in emulator construction.

Realizations from the Mira-Titan simulation suite include 16 lower-resolution (LR) simulations and one higher-resolution (HR) simulation for each cosmology.
The LR simulations are only valid in the linear to mildly non-linear regime, i.e., at small- to medium- values of $k$, and become biased at higher $k$; the HR simulations remain valid all the way into the non-linear regime.
Additionally, a theoretical realization for each cosmology is available that is valid in the linear regime, i.e., at small $k$.
The challenge in learning a set of cosmology-specific smooth matter power spectra from these data is allowing for multiple observations of an underlying nonstationary function, that may cover different regions and have different amounts of noise, to be combined to generate a smooth estimate.
Furthermore, there are common patterns across spectra (e.g., the so-called ``baryonic acoustic oscillation'' (BAO) region, a region of higher wiggliness, appears in the same location of the domain $k$ across different cosmologies), suggesting that sharing information across cosmologies could be useful.

Various approaches to modeling non-stationary processes have been proposed in the literature, including partition models \citep{gramacy2008bayesian}, deep Gaussian processes (GPs) \citep{sauer2023vecchia, salimbeni2017doubly}, and convolution models \citep{sans2008bayesian, higdon2022non}.
\citet{gramacy2008bayesian} proposed partition models that employ treed partitioning to divide the input space into regions and fit separate GPs to each region, thereby accommodating different levels of smoothness. However, this often leads to a function fit with discontinuities.
\citet{sauer2023vecchia, salimbeni2017doubly} introduced deep GPs, which allow for varying degrees of smoothness without the need for partitioning and result in continuous function fits. \citet{higdon2022non} developed a convolutional model that enables the spatial dependence structure to vary according to location. Neither of these methods are able to share information about smoothness patterns across multiple functions.
\citet{sans2008bayesian} proposed a non-stationary model based on discretized convolutions, offering flexibility in capturing space-time covariance structures without relying on pre-specified formulations. 
However, this method requires convolving across the spatial domain to induce non-stationarity in the time domain, and vice versa.
While these existing non-stationary models offer valuable insights, all are limited by their inability to leverage information sharing across related but distinct functions.
DPC overcomes these limitations. 
It produces a continuous functional fit that accommodates different levels of smoothness and is able to share information about smoothness patterns across multiple functions.


Although the DPC method has been briefly described in the Coyote and Mira-Titan emulation papers, the method has not been rigorously described or tested, and no code has been released.
Here we describe the DPC method, assess its coverage, compare its performance to competitors using simulated data, show results of using it to learn the power spectra in the Mira-Titan simulation suite, and release a public R package for using the method.

\section{Model}
\subsection{Process Convolutions}
A stochastic process convolution can be thought of as a type of moving average of another stochastic process.  Formally, the convolution process $z(s)$ over a general spatial region $\mathcal{S}$ is created from the stochastic process $y(\cdot)$ as 

\begin{equation} \label{eq:procconv}
z(s) = \int_{\mathcal{S}} k(x-s)y(x)dx,\ \text{for } s \in \mathcal{S}.
\end{equation}

Here, $k(\cdot)$ is a kernel function (e.g. the normal probability density function) which governs the smoothness of $z(s)$ for a given process $y(\cdot)$.  In practice, however, the process $y(\cdot)$ is often restricted to lie on a finite subset of $n_x$ points $\mathcal{S}_x = \{x,\ldots,x_{n_x}  \} \subset \mathcal{S}$ to reduce the computational demand.  Thus \eqref{eq:procconv} simplifies to

\begin{eqnarray}
z(s) &=& \int_{\mathcal{S}_x} k(x-s)y(x)dx,\ \text{for } s \in \mathcal{S} \nonumber \\
&=& \sum_{i=1}^{n_x} k(x_i-s)y(x_i). \label{eq:convsum}
\end{eqnarray}

\begin{figure}
\begin{center}
\includegraphics[width=0.7\textwidth]{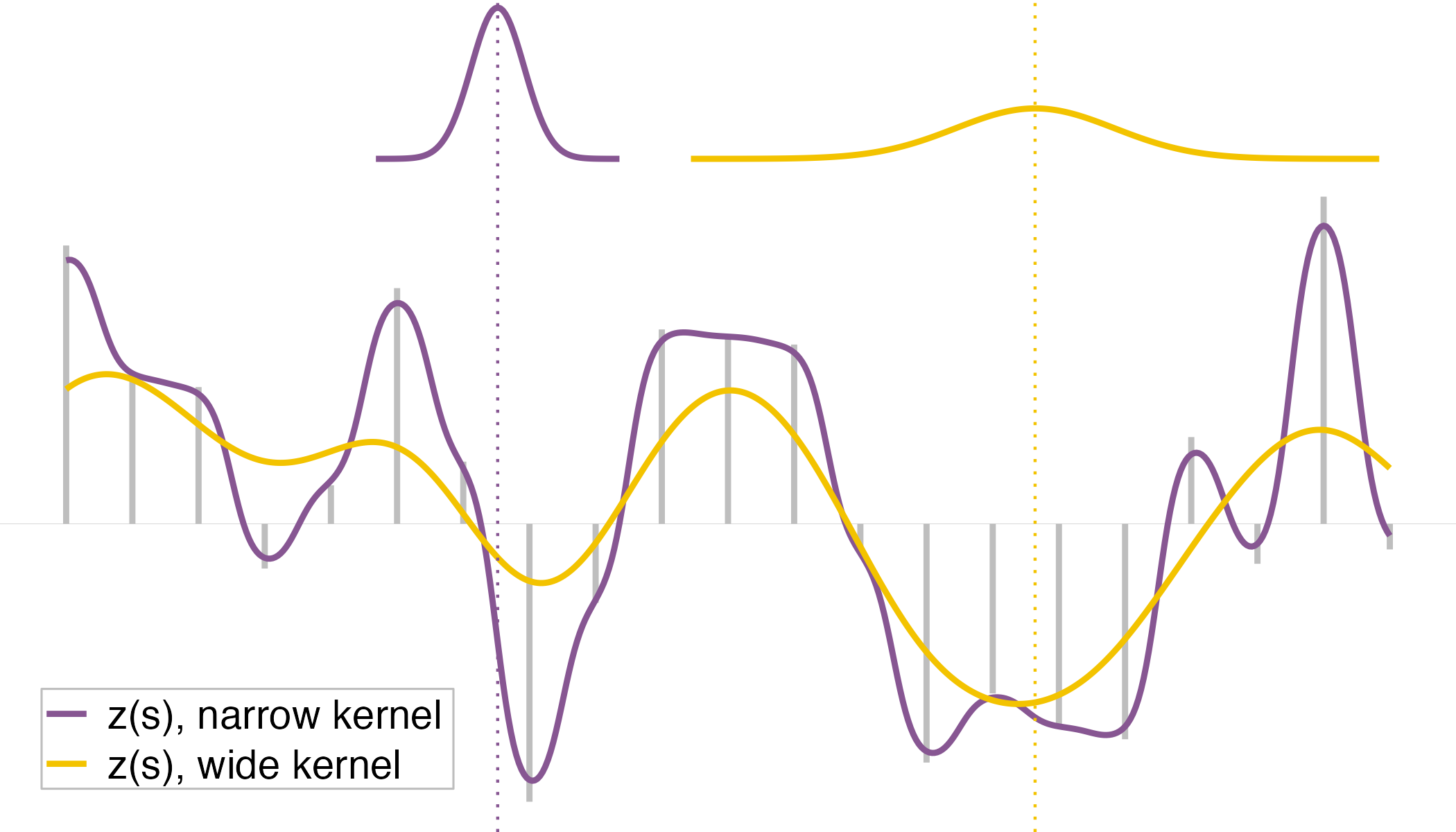}
\caption{One realization of the process $y(\cdot)$ (vertical lines) and the induced realization of the process convolution $z(\cdot)$ using two different normal kernels with standard deviations of 0.3 (wide) and 0.1 (narrow). The kernels, shown above, illustrate how a given point on $z(\cdot)$ is the weighted average of the process $y(\cdot)$ with weights proportional to the kernel value at each location. Since the wider kernel averages over a larger effective set of process values, the resulting function is fairly smooth, whereas the function induced by the narrower kernel is more wiggly.}
\label{fig:conv}
\end{center}
\end{figure}

As an example, suppose we define $z(s)$ over $\mathcal{S} = [0,5] \subset \mathbb{R}$ and we set the locations for the process $y(\cdot)$ to be $\mathcal{S}_x = \{x,\ldots,x_{n_x}  \} =  \{0,0.25,0.5,\ldots4.75,,5\}$. Further suppose that each $y(x_i)$ are independently and identically distributed $N(0,1)$.  Figure~\ref{fig:conv} shows one realization of the process $y(\cdot)$ and the induced realization of the convolution process $z(\cdot)$ with two different kernels over the domain $\mathcal{S}$.  The vertical lines at $\mathcal{S}_x$ are the realizations of $y(\cdot)$.  The overlaid smooth purple and yellow lines represent the induced process $z(\cdot)$ with a normal kernel with standard deviations 0.1 and 0.3, respectively; these kernels are shown above the process realizations, with vertical dashed lines marking the location at which the realization is being calculated using that kernel.  
Notice how the realization of $z(\cdot)$ is less smooth using a kernel with a standard deviation of 0.1 and is more smooth using a kernel with a larger standard deviation of 0.3.

One of the main advantages to using process convolutions is the fact that one does not need to directly specify the covariance function of the process convolution $z(\cdot)$.  Although there are some one-to-one correspondence between certain process convolutions and standard covariance functions, process convolutions provide a way to easily construct processes with a variety of desirable properties.  For instance, we assume a Brownian motion process for $y(\cdot)$ to provide more stable tail behavior when modeling cosmology data, but we do not need to explicitly specify the covariance function of $z(\cdot)$.

When using process convolutions to model data, we generally assume that we have observed the process convolution, but not the process from which it was induced.  Therefore, the process which is convolved to create a process convolution ($y(\cdot)$ in the example above) is considered a latent function for modeling purposes. Thus, given a kernel family, interest lies in estimating the latent function and any relevant parameters of the kernel function to obtain an estimate of the convolved process $z(\cdot)$.

\subsection{Bayesian Model}
Let $P_{ij}$ be the $j$th realization from the $i$th process whose mean is $\Ps_i$.  For the Mira data, $P_{ij}$ corresponds to the $j$th replication (simulation) of the $i$th cosmology/time combination. Specifically, we assume a multivariate-normal distribution on $P_{ij}$,
\[
f(P_{ij}) \propto |\Omega_{ij}|^{\frac{1}{2}} \exp\left\{-\frac{1}{2}(P_{ij} - A_{ij}\Ps_i)^T \Omega_{ij} (P_{ij} - A_{ij}\Ps_i) \right\},
\]
and assume that $\Omega_{ij}$ is a known precision matrix.  In the present model, we allow each  process realization, $P_{ij}$, to occur at any number of locations and at any location (i.e. we do not assume that realizations have the same number of locations nor occur on a subset of the same grid).  To accommodate this feature of the model, the projection matrix $A_{ij}$ subsets $\Ps_i$ to the locations of the realized process $P_{ij}$ since evaluating the likelihood requires $\Ps_i$ to be known only at the locations of the realized $P_{ij}$.  

The process convolutions are used to model the smooth mean functions $\Ps_i$, $i=1,\ldots,n$.  We model $\Ps_i$ using a process convolution prior on Brownian motion.  Specifically, $\Ps_i = K^\sigma u_i$ where the $K^\sigma$ is an $N \times m$ convolution matrix with elements
\[
K^\sigma_{i^\prime,j^\prime} = \frac{1}{\sqrt{2\pi\sigma_{i^\prime}^2}} \exp\left\{ \frac{(k_{i^\prime} - x_{j^\prime})^2}{2\sigma_{i^\prime}^2} \right\}.
\]
Here, $k_1,\ldots,k_N$ represent the $N$ unique locations for all realizations $P_{ij}$ and $u_i$ is a latent Brownian motion vector which is realized on a sparse grid of $m$ (usually) evenly spaced points, $x = (x,\ldots,x_m)$.  The prior density on $u_i$ is 
\[
f(u_i) \propto \left|\frac{1}{\tau^2_u} W \right|^\frac{1}{2} \exp\left\{-\frac{u_i^\prime W u_i}{2\tau^2_u} \right\}.
\]
where $W$ is the Brownian motion precision matrix with diagonal equal to $[1\ 2\ \cdots\ 2\ 1]$ and -1 on the first off-diagonals.  Note that each element of $\Ps_i = K^\sigma u_i$ is similar to \eqref{eq:convsum} except: 1) here we are using Brownian motion instead of i.i.d.\ Gaussian variates for the latent process and 2) the normal kernel bandwidth parameter changes over the domain.

In order to allow the bandwidth parameter, $\sigma$, to vary over the domain we place a second process convolution model on $\sigma$.  This second process convolution layer is what makes this model a so-called ``deep" process convolution and allows us to model data which is non-stationary. Specifically, $\sigma = K^\delta v$ where the elements of the $N \times n_v$ matrix $K^\delta$ are
\[
K^\delta_{i^\star,j^\star} = \frac{1}{\sqrt{2\pi\delta^2}} \exp\left\{-\frac{(x_{i^\star} - t_{j^\star})^2}{2\delta^2}\right\}.
\]
Here, $t_j$, $j=1,\ldots,n_v$ represent the locations of the realizations of the $n_v$-length vector $v$.  These locations are placed on a sparser grid than the $u$'s. This second process on the bandwidth parameter is built on i.i.d. Gaussian variates, 
\[
f(v) \propto \left(\frac{1}{\sqrt{\tau^2_v}}\right)^{n_v} \exp\left\{- \frac{v^\prime v}{2\tau^2_v}\right\}
\]
with the necessary restriction that all the elements of $\sigma = K^\delta v$ are positive. 

Lastly, vague priors are placed on $\tau^2_u,\ \tau^2_v$, and $\delta$ yielding
\begin{eqnarray*}
\pi(\tau^2_u) &\propto& \left(\tau^2_u\right)^{-\alpha_u - 1} \exp\left(- \frac{\beta_u}{\tau^2_u}\right), \\ 
\pi(\tau^2_v) &\propto& \left(\tau^2_v\right)^{-\alpha_v - 1} \exp\left(- \frac{\beta_v}{\tau^2_v}\right), \\ 
\pi(\delta) &\propto& I\{0 \leq \delta \leq \delta_{\max}\},
\end{eqnarray*}
which are inverse-gamma, inverse-gamma, and uniform, respectively.  We set the default prior hyperparameters to $\alpha_u = \alpha_v = 1$, $\beta_u = \beta_v = .001$, and $\delta_{\max}$ such that it has large enough support (e.g. 10, but this may need to be larger for datasets collected over a larger spatial domain). 
We can now write the entire posterior as 
\begin{eqnarray*}
\pi\left(\{u_i\}_{i=1}^n,v,\tau^2_u,\tau^2_v,\delta \ \left| \   \{P_{ij}\}_{\substack{\text{$i=1,\ldots,n$} \\ \text{$j=1,\ldots,n_i$}}}  \right. \right) \propto \\ 
\prod_{i,j} |\Omega_{ij}|^{\frac{1}{2}} \exp\left\{-\frac{1}{2}(P_{ij} - A_{ij}K^\sigma u_i)^T \Omega_{ij} (P_{ij} - A_{ij}K^\sigma u_i) \right\} \times \\
\prod_i \left|\frac{1}{\tau^2_u} W \right|^\frac{1}{2} \exp\left\{-\frac{u_i^\prime W u_i}{2\tau^2_u} \right\} \times \\
\left(\frac{1}{\sqrt{\tau^2_v}}\right)^{n_v} \exp\left\{- \frac{v^\prime v}{2\tau^2_v}\right\} \times \\
\left(\tau^2_u\right)^{-\alpha_u -1} \exp\left(- \frac{\beta_u}{\tau^2_u}\right) \times \\
\left(\tau^2_v\right)^{-\alpha_v - 1} \exp\left(- \frac{\beta_v}{\tau^2_v}\right) \times \\ 
I\{0 \leq \delta \leq \delta_{\max}\}.
\end{eqnarray*}

Since the posterior does not have an analytical form, Markov chain Monte Carlo (MCMC) methods must be used to obtain draws from the posterior.  To reduce the dimensionality of the parameter space in the MCMC algorithm, the latent processes $u_i$, $i=1,\ldots,n$ can be integrated out, yielding a more efficient MCMC algorithm on the marginal posterior of $v,\ \tau_u^2,\ \tau_v^2,$ and $\delta$.

\begin{lem}
	\label{lem:u}
	The marginal posterior for $v,\ \tau^2_u,\ \tau^2_v$, and $\delta$ can be expressed as 
	
	\begin{eqnarray*}
		\pi\left(v,\tau^2_u,\tau^2_v,\delta \ \left| \   \{P_{ij}\}_{\substack{\text{$i=1,\ldots,n$} \\ \text{$j=1,\ldots,n_i$}}}  \right. \right) \propto \\ 
		\prod_{i=1}^n \left[ \exp\left\{ \frac{1}{2} \left(\sum_{j=1}^{n_i} D_{ij} \right)^T C_i^{-1} \left(\sum_{j=1}^{n_i} D_{ij} \right) \right\} \times  \left| C_i^{-1} \right|^{\frac{1}{2}} \right] \times \\
		\left(\frac{1}{\tau^2_u}\right)^{\frac{n n_u}{2}} \times 
		\left(\frac{1}{\tau^2_v}\right)^{\frac{n_v}{2}} \exp\left\{- \frac{v^\prime v}{2\tau^2_v}\right\} \times \\ \left(\tau^2_u\right)^{-\alpha_u -1} \exp\left(- \frac{\beta_u}{\tau^2_u}\right) \times 
		\left(\tau^2_v\right)^{-\alpha_v - 1} \exp\left(- \frac{\beta_v}{\tau^2_v}\right) \times \\ 
		I\{0 \leq \delta \leq \delta_{\max}\},
	\end{eqnarray*}

where $B_{ij} = {K^\sigma}^T A_{ij}^T\Omega_{ij}A_{ij}K^\sigma$, $C_i = \sum_{j=1}^{n_i} B_{ij} + \frac{W}{\tau^2_u}$, and $D_{ij} = {K^\sigma}^T A_{ij}^T \Omega_{ij} P_{ij}$.
\end{lem}

From the marginal posterior derived in Lemma~\ref{lem:u}, we obtain the posterior of $\tau^2_v$ conditional on all other parameters (with the vector $u$ marginalized),
\begin{eqnarray*}
	\pi(\tau^2_v \mid \cdot) \propto \left(\frac{1}{\tau^2_v}\right)^{\frac{n_v}{2}} \exp\left\{- \frac{v^\prime v}{2\tau^2_v}\right\} \left(\tau^2_v\right)^{-\alpha_v - 1} \exp\left(- \frac{\beta_v}{\tau^2_v}\right) \\
	= \left(\tau_v^2 \right)^{-\alpha_v - 1 - \frac{n_v}{2}} \exp\left\{ -\frac{1}{\tau^2_v}\left(\frac{v^\prime v}{2} + \beta_v \right)\right\}
\end{eqnarray*}
which implies $\pi(\tau^2_v \mid \cdot) \sim IG(\alpha_v + n_v/2,\ v^\prime v /2 + \beta_v)$.

The proof of Lemma~\ref{lem:u} also reveals the conditional posterior of $u_i$, which is useful in drawing values of $u_i$ to construct credible intervals for the mean $\Ps_i$.  Specifically, 
\begin{eqnarray*}
	\pi\left(u_i | \cdot \right) \propto
	\prod_{j=1}^{n_i} \exp\left\{-\frac{1}{2}(P_{ij} - A_{ij}K^\sigma u_i)^T \Omega_{ij} (P_{ij} - A_{ij}K^\sigma u_i) \right\} \times 
	\exp\left\{-\frac{u_i^\prime W u_i}{2\tau^2_u} \right\} \\
	\propto (2\pi)^{- \frac{1}{2}n_i} \left| C_i^{-1} \right|^{-\frac{1}{2}} \exp\left\{ -\frac{1}{2}  \left[\left(u_i - C_i^{-1} \sum_{j=1}^{n_i} D_{ij} \right)^T C_i \left(u_i - C_i^{-1}\sum_{j=1}^{n_i} D_{ij} \right)  \right]\right\}
\end{eqnarray*}
which implies $u_i \mid \cdot \sim MVN(C_i^{-1}\sum_{j=1}^{n_i} D_{ij}, C_i^{-1})$ where $C_i$ and $D_{ij}$ are the same as in Lemma~\ref{lem:u}.

The MCMC algorithm proceeds by sequentially updating $\tau_u^2,\ \delta$, and the elements of $v$ using Metropolis Hasting (MH) steps and updating $\tau_v^2$ via a Gibbs step.  Posterior draws of each mean function $\Ps_i$ can be obtained by constructing $K^\sigma$ and drawing $u_i$ for each posterior draw and computing  $K^\sigma u_i$.

\section{Data}

In this section, we describe two types of datasets: synthetic data generated from known underlying functions, and the Mira cosmological dataset describing the power spectra learned in large N-body simulations. The synthetic datasets are employed to benchmark our method against competitors, offering a controlled setting to evaluate performance. The Mira dataset, on the other hand, is used to illustrate the practical application of our method, emphasizing its effectiveness and usefulness for analysis problems relying on multi-resolution nonstationary data.

\subsection{Simulated Data}\label{sec:simfuncs}



\begin{figure}
	\begin{center}
    \includegraphics[width=0.45\textwidth]{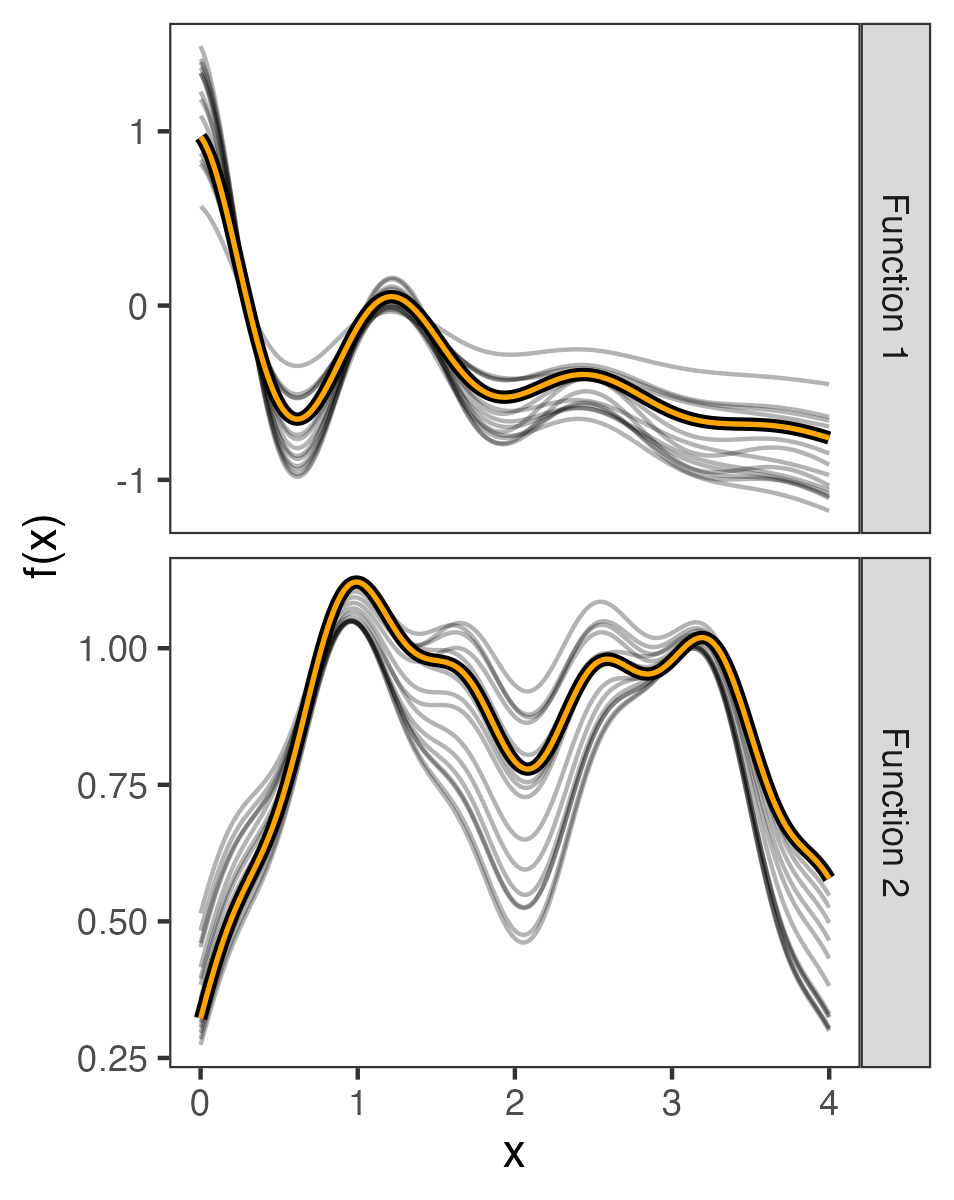}
	\caption{Draws of noise-free functions $f_1$ and $f_2$ under different $m$ and $u$ samples; the highlighted samples in each sub-plot are shown again in Figure \ref{fig:simex_noisy}.}
	\label{fig:simex_funcs}
	\end{center}
\end{figure}

We assess how well the model does at learning arbitrary smooth functions using simulated data. We simulate data using two analytic functions, $f_1$ and $f_2$, having varying levels of smoothness across their domains. The functions are specified as follows:
\begin{equation}
\begin{split} 
    f_1(x) &= m_1\exp(-u_1 \mkern 1mu x/2) * \cos(w \mkern 1mu x) - m_1 \mkern 1mu x/5, \ w = \sqrt{25 - [u_1/2]^2}, \\
    f_2(x) &= \exp(-m_2[x-3]^2) + \exp(-u_2[x-1]^2) - 0.05 \mkern 1mu  \sin(8 \mkern 1mu [x-1.9]), \\ x &\in [0,4].
\end{split}
\end{equation}
To allow for structurally related but distinct functions, we sample rather than fix $m_1, m_2, u_1,$ and $u_2$. Specifically, we sample $m_1 \sim \text{Uniform}(0.5, 1.5), u_1 \sim \text{Uniform}(1.5, 2.5),$ and $m_2, u_2 \sim \text{Uniform}(0.6, 1.4)$. Resulting draws from $f_1$ and $f_2$ are shown in Figure \ref{fig:simex_funcs}. 

For our simulation experiment we sample noisy realizations from each underlying true function. Specifically, we have
\begin{equation}
\begin{split}
    y_1(x) &= f_1(x) + \epsilon_1(x), \ \epsilon_1 \sim \text{Normal}(0,\sigma_1(x)^2), \\
    y_2(x) &= f_2(x) + \epsilon_2(x), \ \epsilon_2 \sim \text{Normal}(0,\sigma_2(x)^2).
\end{split}
\end{equation}
We consider three noise ``settings''. 
\begin{itemize}
\item Setting A: homoskedastic noise. Function standard deviations are $\sigma_1(x)=0.01,$ and $\sigma_2(x)=0.015.$
\item Setting B: heteroskedastic noise defined by a continuous function. Function standard deviations are $\sigma_1(x)=\frac{1}{10[x+1]}$ and $\sigma_2(x)=\frac{\sqrt{|x-2|}}{10}.$ 
\item Setting C: heteroskedastic noise defined by a piecewise continuous function. Function standard deviations are $10^{-7}$ for the left-most `piece' (i.e., $x \leq 0.4$ for $f_1$ and $x \leq 0.5$ for $f_2$), then become a scaled version of the Setting B standard deviations with a scaling factor of 0.75 for the middle `piece' (i.e., $0.4 < x < 3$ for $f_1$ and $0.5 < x < 2$ for $f_2$) and 0.5 for the right-most `piece' (i.e., $x \geq 3$ for $f_1$ and $x \geq 2$ for $f_2$). The left-most near-0 variance term is meant to mimic the existence of a theoretically known segment of the function.
\end{itemize}
Further sampling details are included in the Supplemental Materials. 
The Supplemental Materials also includes a consistency study using data simulated from the DPC model; we show that the credible intervals are appropriate and the parameter posteriors converge to their true values as $N$ increases.

\begin{figure}
	\begin{center}
    \includegraphics[width=0.9\textwidth]{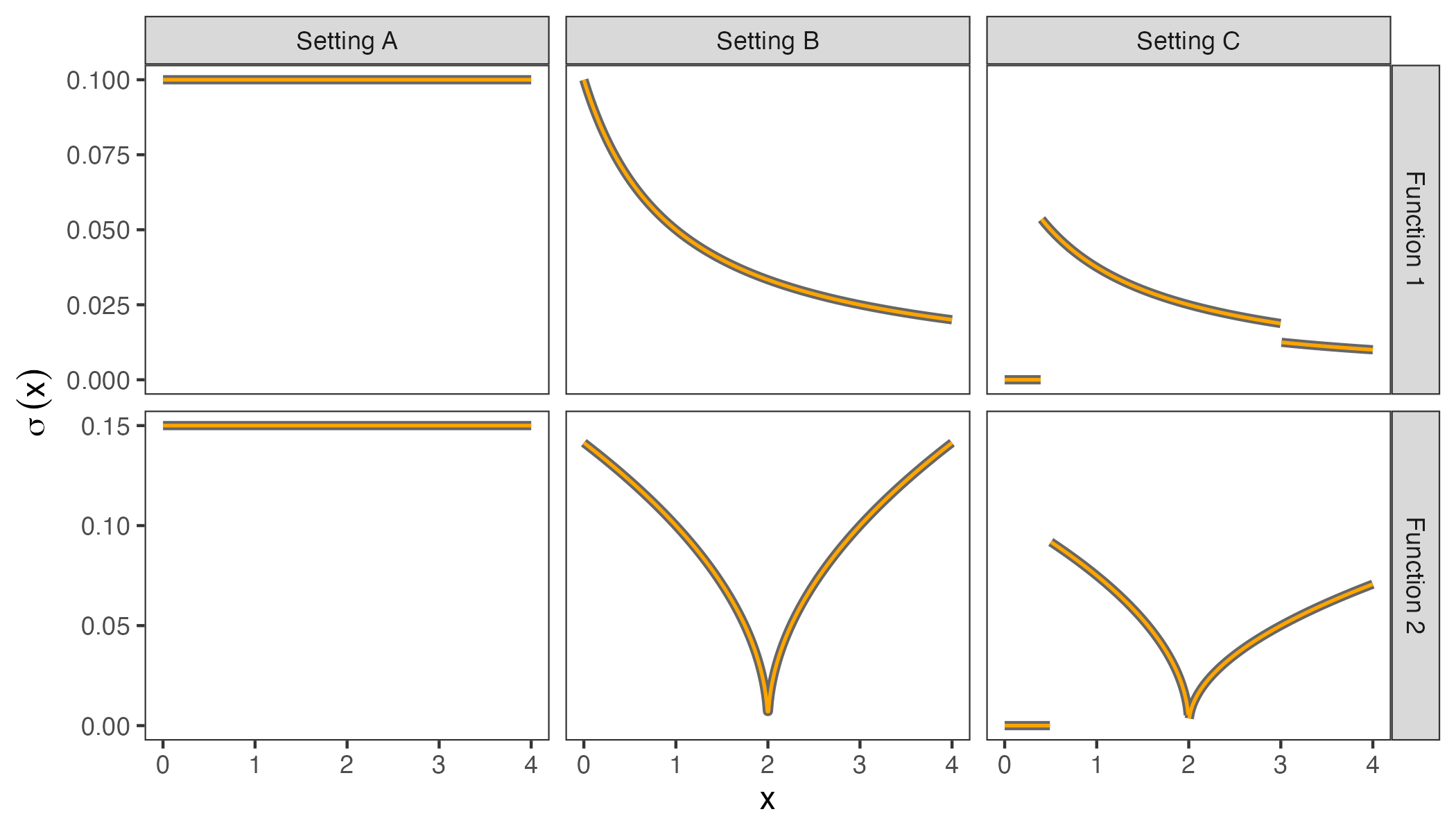}
	\caption{Standard deviation values by input $x$ for each of the simulation settings. Setting A uses homoskedastic variance, while Settings B and C use heteroskedastic variance. Setting C includes an early part of the domain in which the variance is near zero.}
	\label{fig:simex_var}
	\end{center}
\end{figure}

\begin{figure}
	\begin{center}
    \includegraphics[width=0.9\textwidth]{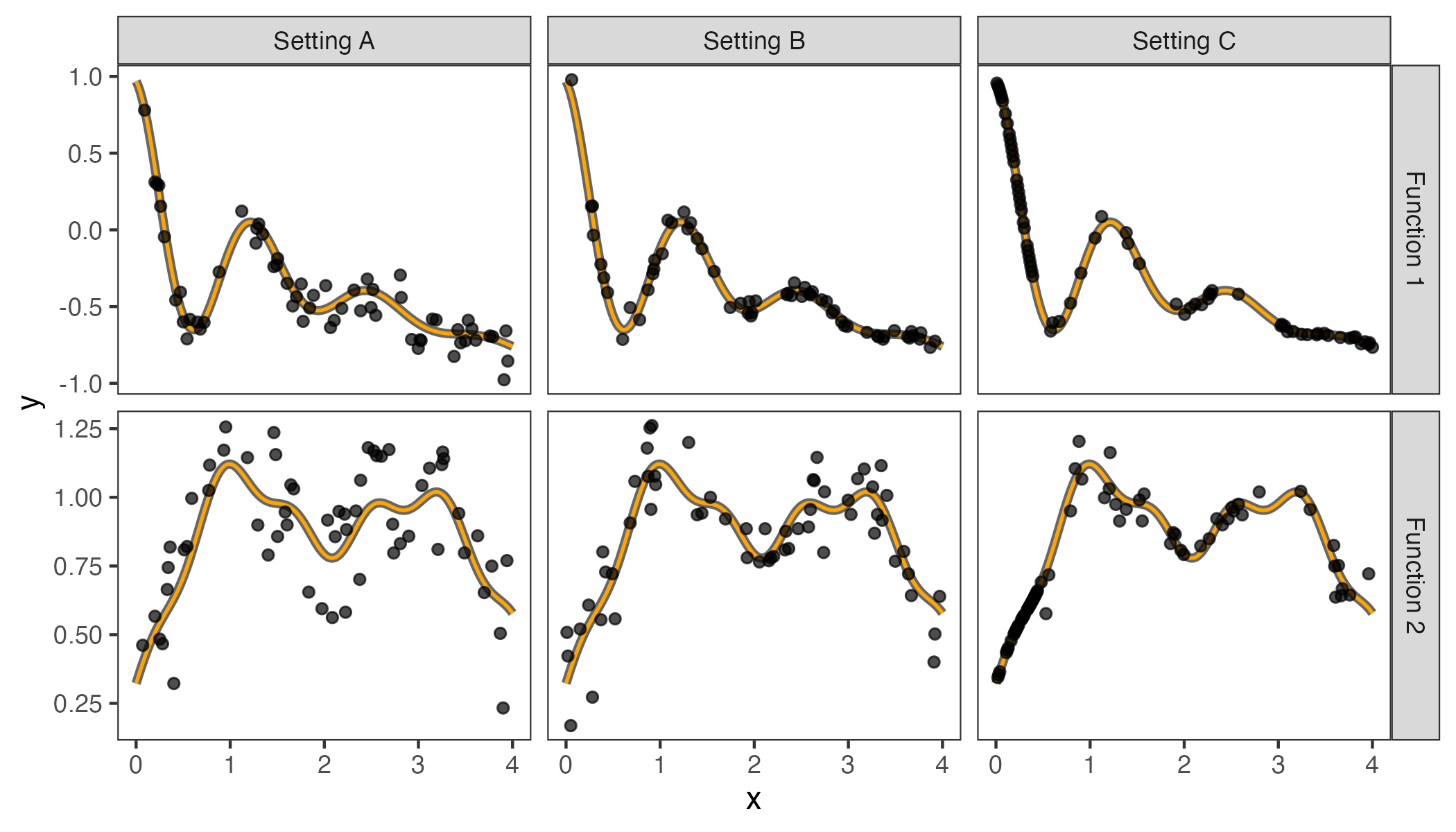}
	\caption{Example noisy realizations from each setting of two underlying true functions (specifically, the highlighted curves from Figure \ref{fig:simex_funcs}), where the variance of the draws is that shown in Figure \ref{fig:simex_var}.}
	\label{fig:simex_noisy}
	\end{center}
\end{figure}

\subsection{Cosmological Data}

The Mira-Titan Universe simulation suite is a collection of gravity-only simulations designed to investigate the six standard cosmological parameters: ${\omega_m,\omega_b, \sigma_8, h, n_s, w_0}$, as well as the effects of massive neutrinos and dynamic dark energy, represented by parameters ${\omega_{\nu}, w_a}$.
The simulations cover 111 cosmologies at a redshift (i.e., stage of the universe's expansion) range of $z=0$ to $z=2.02$ with a maximum wave number of $k=5$~Mpc$^{-1}$. 
For each cosmology, the suite includes one high-resolution (HR) simulation at each redshift, each modeling a volume of (2.1 Gpc)$^3$ using 3200$^3$ particles. 
Additionally, the suite includes 1776 lower-resolution (LR) simulations (16 per cosmology and red shift combination), and the result from Time-Renormalization Group perturbation theory (PT) for each cosmology and red shift. 
The suite was first introduced by \cite{heitmann2016mira}. 
Detailed information on the design and implementation of the simulations can be found in \cite{lawrence2017mira}. 
All simulations utilized the Hardware/Hybrid Accelerated Cosmology Code (HACC), running on CPU (Mira) and CPU/GPU (Titan) platforms with optimized force computation algorithms for each setup.

\begin{figure}
\begin{center}
\includegraphics[width=0.8\textwidth]{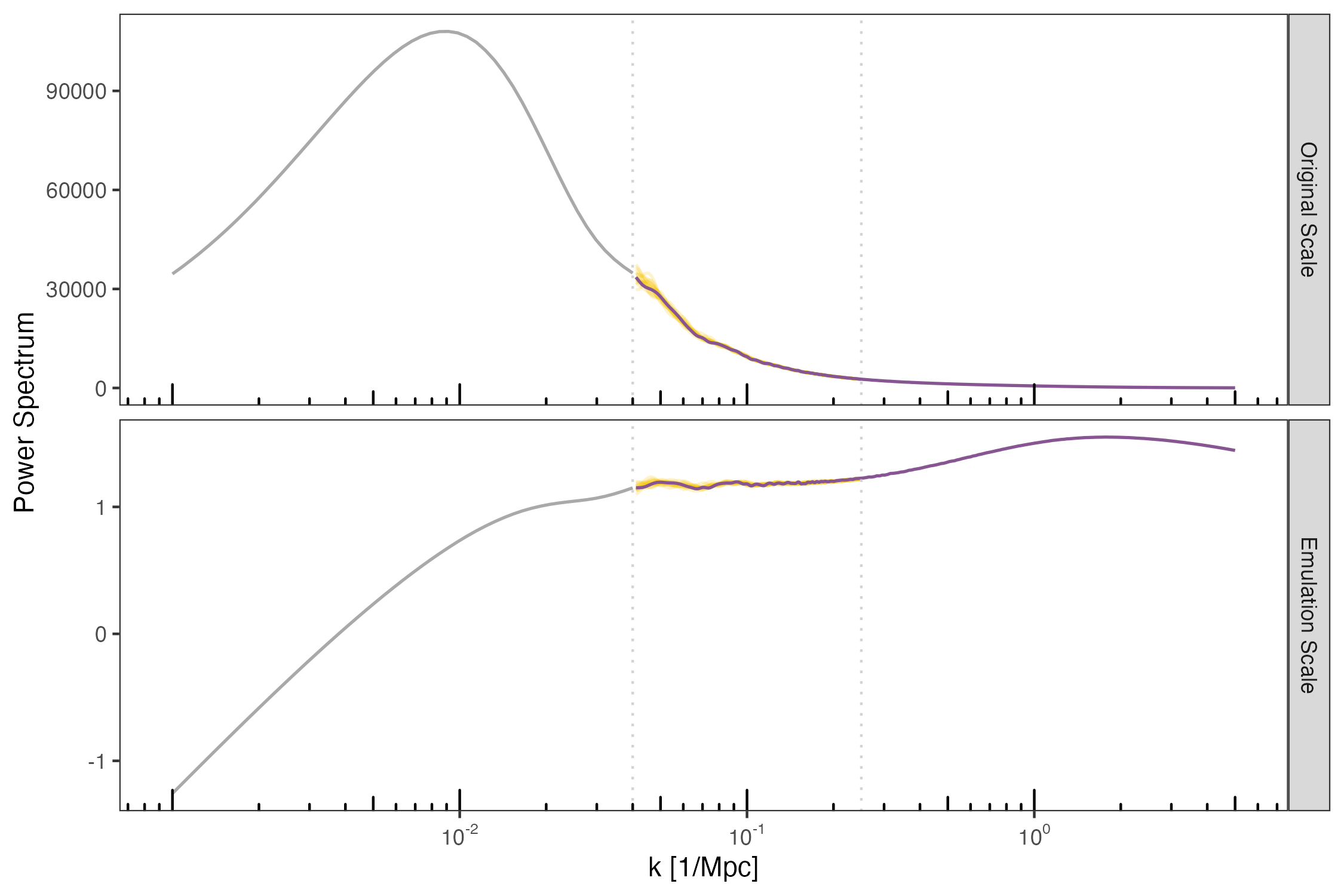}
\caption{Data from an example cosmology (M098, as described in \cite{moran2023mira}) shown at redshift $z=0$. The left vertical dashed line denotes where we stop using PT and begin using LR and HR data, and the right line marks where we stop using LR runs; these cutoffs were specified by domain scientists.}
\label{fig:data_mtnoisy}
\end{center}
\end{figure}

Figure \ref{fig:data_mtnoisy} shows the power spectrum realizations $P(k)$ for a single cosmology and red shift on the original scale and the so-called ``emulation scale'' that accentuated the BAO region, where $\mathcal{P}(k) = \log_{10}\big({k^{1.5} P(k)}/{2\pi^2}\big) = \log_{10}(\Delta^2(k) / k^{1.5})$ denotes the transformed spectrum.
The data exhibit varying smoothness over the domain, with the BAO region, in which both LR and HR data are used, being more wiggly.
It is known that as $k$ gets larger, each of the PT, LR, and HR realizations have increased bias to differing severities. 
Subject matter experts determine appropriate cutoffs to maintain the integrity of the data.
The PT realizations were kept for values of $k$ below 0.04., LR runs at $k$ values less than 0.25, and HR runs at values of $k$ below 5.
In regions where PT theory is reliable, no LR or HR runs are needed because the PT is considered to be a precisely known observation of the underlying true spectrum.
After implementing these cutoffs, a total of 340,587 observations were used in the analysis.

The deep process convolution model assumes that the variances are known for each data point.  
In our case, the variance for an individual data point can be computed theoretically for these simulations, but these calculations yield noisy estimates of the variances. 
We pool information across cosmologies to learn a variance value rather than using the noisy estimates.  
We first transform the theoretical variances computed for the spectra on the original scale under the assumption that the original data are log-normal.  
An approximation of the variance on the log scale can then be computed by solving $\kappa^2 = (\exp(\sigma^2) - 1) \exp(2\mu + \sigma^2)$ for $\sigma^2$ and estimating $\mu$ using the mean of the transformed data.  
This yields $\hat{\sigma}^2 = \log[.5(1 + \sqrt{1 + 4\exp(\log(\kappa^2) - 2\hat{\mu})})]$ where $\kappa^2$ is the theoretical variance of a data point on the original scale, and $\hat{\mu}$ is the estimated mean of the transformed data at $k$, the location of the simulated spectra value.  
For LR data, $\hat{\mu}$ was computed using the sample mean of all of the low-res runs at the value $k$.  
For HR runs (where there was only one replication), the value of the high-resolution run at the value of $k$ was used.

\section{Results}
\subsection{Simulations}







The performance metrics we consider in estimating the underlying true functions described in Section \ref{sec:simfuncs} are the mean squared error (MSE), the coverage of the 95\% uncertainty interval, and the 95\% uncertainty interval width. 
These metrics are evaluated over a grid of 400 points spaced evenly along the domain.

\begin{figure}
	\begin{center}
    \includegraphics[width=0.7\textwidth]{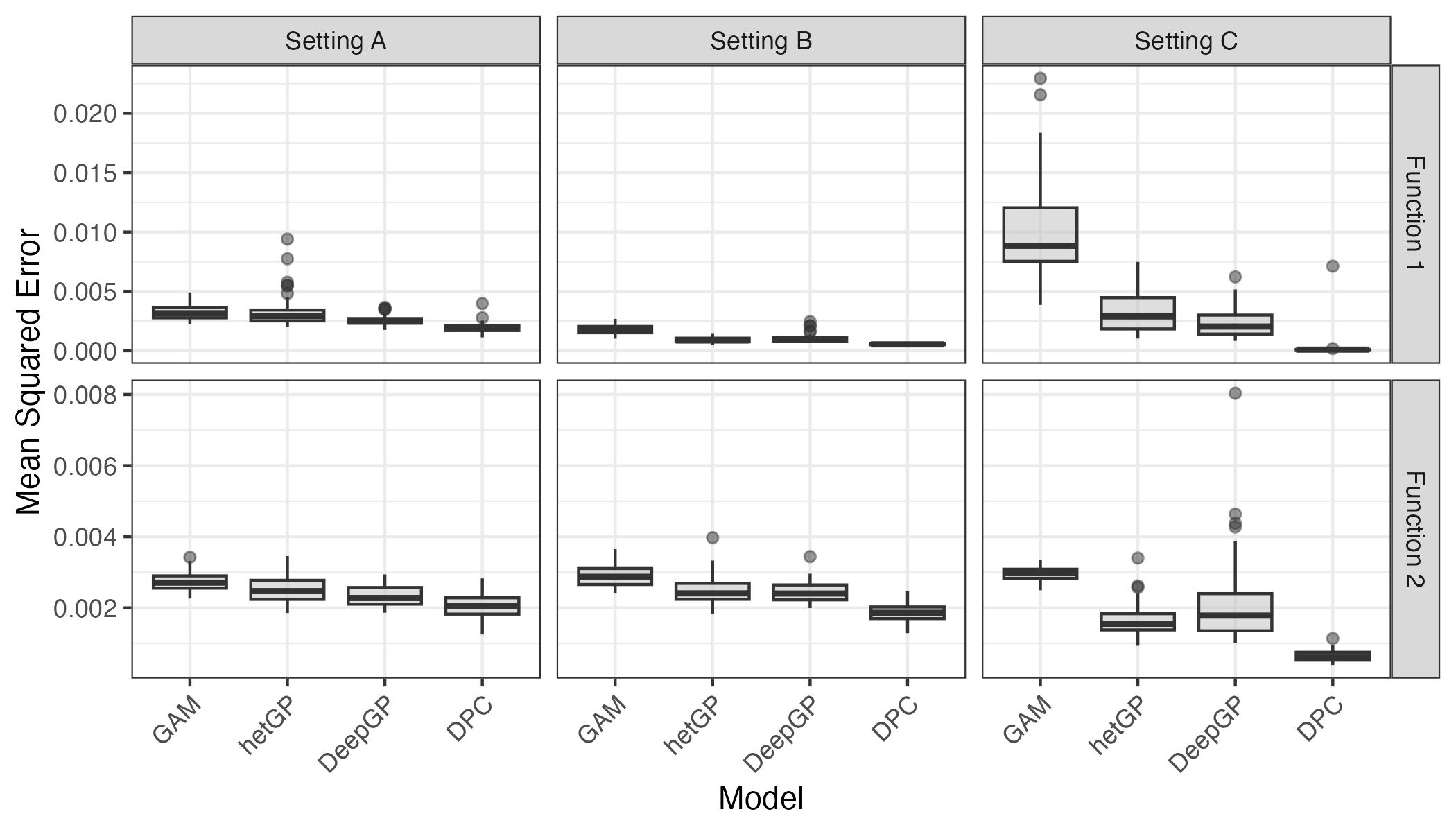}
	\caption{MSE of estimating the underlying true functions for the 50 replicate simulations performed under each function and setting. Lower means better average fit between estimated and true functions.}
	\label{fig:sim_mse_zoom}
	\end{center}
\end{figure}

Figure \ref{fig:sim_mse_zoom} shows box plots summarizing the MSE under the 50 replicate simulations for estimating the functions using each model, by function type and setting.
The DPC model has the lowest replicate-averaged MSE under each function/setting combination.
Furthermore, it tends to have a better ``worst-case'' than competitors; i.e., the max MSE across replicates tends to be lower for the DPC than the other models (en exception being Function 1 under Setting C, in which the DeepGP has a lower worst-case MSE).
If we look at a replicate-level head to head comparison, the DPC has the lowest MSE in 97.7\% of simulation replicates.

\begin{figure}
	\begin{center}
    \includegraphics[width=0.7\textwidth]{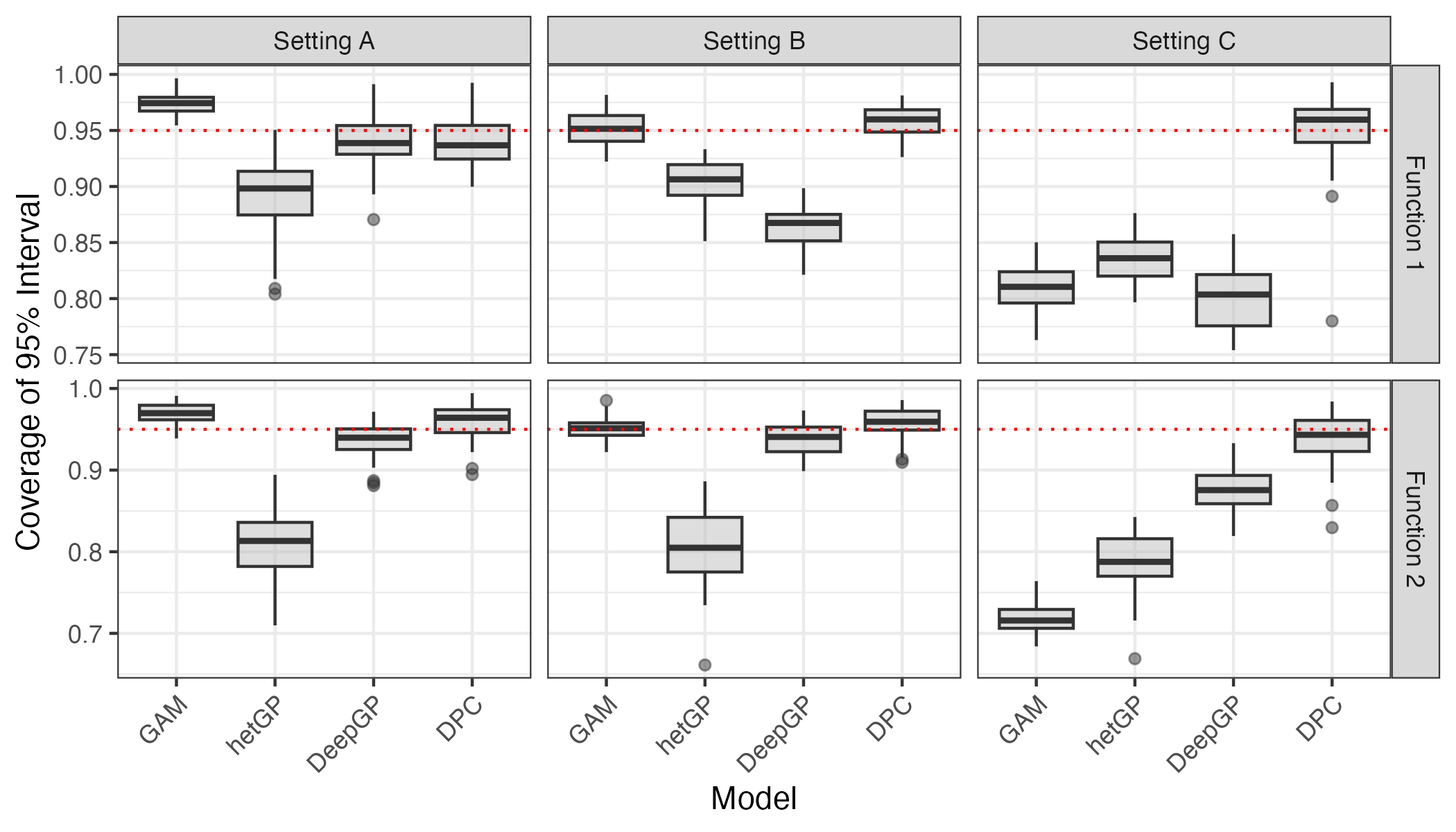}
	\caption{Coverage of the 95\% credible intervals for estimating the underlying smooth function in the 50 replicate simulations performed under each function and setting. Closer to 95\% coverage (denoted by the horizontal red dashed line) indicates more fidelitous uncertainty intervals.}
	\label{fig:sim_cvg}
	\end{center}
\end{figure}

\begin{figure}
	\begin{center}
    \includegraphics[width=0.7\textwidth]{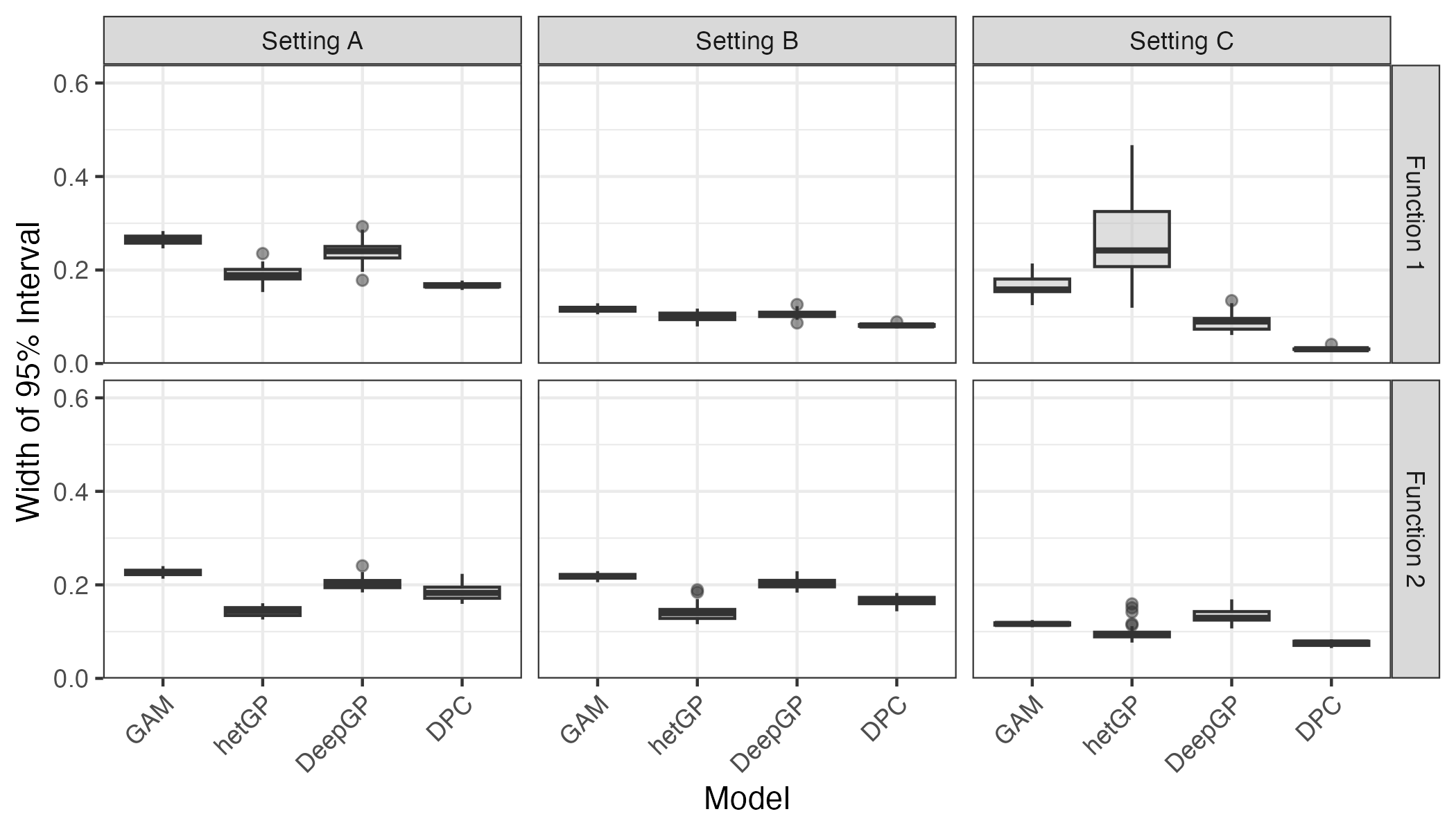}
	\caption{Average 95\% credible interval widths for the function estimate in the 50 replicate simulations performed under each function and setting. Lower means a more precise estimate of the functions. For visual clarity one outlier in the hetGP model for Setting C of Function 1 has been omitted from the plot.}
	\label{fig:sim_wid}
	\end{center}
\end{figure}

Figures \ref{fig:sim_cvg} and \ref{fig:sim_wid} show box plots summarizing the the coverage of the 95\% uncertainty intervals and  the average 95\% uncertainty interval widths, respectively, under the 50 replicate simulations.
Of the models, the DPC model overall has the best combination of precision and closeness of coverage to nominal.
The hetGP interval widths are generally comparable to the DPC interval widths and narrower than those of the GAM and deepGP models; however, the coverage of these hetGP intervals tend to under-cover relative to nominal coverage on average.
The GAM and DeepGP models produce larger uncertainty intervals than the hetGP or DPC models, and are inconsistent in their coverage.
The DPC model has very narrow credible intervals and overall tends to cover close to nominally, with an average coverage of 95.1\% across setting and function (94.9\% for $f_1$, and 95.2\% for $f_2$).
It is also more consistent in its coverage; note, e.g. that the other models tend toward under- or over-coverage with variation across setting and function.

For these simulations, there is no ``true'' underlying DPC model and so the number and locations of the process must be chosen so as to allow for a ``good enough'' fit of the underlying smooth functions.
Typically, choosing process locations to lie on an evenly spaced grid is a reasonable choice.
Once the number of processes is sufficient to fit the data, adding more locations should not significantly change the learned function.
That is, whether the number of processes is sufficient can be assessed by checking for convergence of function fit, in addition to the traditional residual diagnostics.
In general, the more complicated the function, the more gridpoints may be necessary to adequately model it.
The Supplemental Materials provides an example of the learned function under different numbers of grid points, discusses the associated residual diagnostics for each, and shows how the shape stabilizes as grid density increases.  
The Supplemental Materials also shows how the models perform with different numbers of functions and numbers of observed points per function; as expected, the fewer functions that are observed (i.e., the less shared information to leverage) the less advantage the DPC model enjoys. 

\subsection{Smooth Matter Power Spectra}

\begin{figure}
\begin{center}
\includegraphics[width=0.9\textwidth]{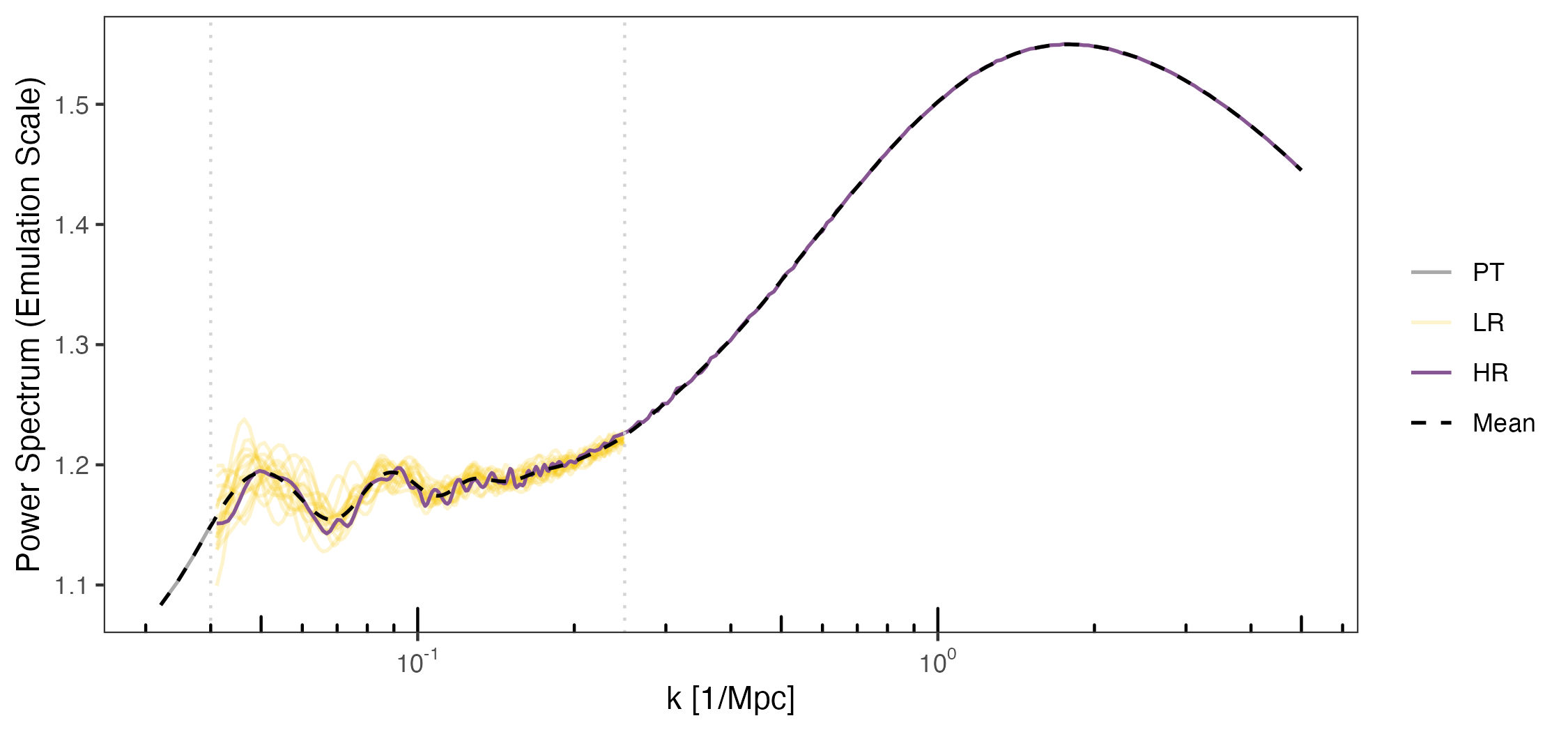}
\caption{Data and learned mean MPS for the cosmology shown in Figure \ref{fig:data_mtnoisy}, with scale zoomed in to emphasize the LR and HR regions.}
\label{fig:data_mtnoisy_wmean}
\end{center}
\end{figure}

\begin{figure}
	\begin{center}
    \includegraphics[width=0.9\textwidth]{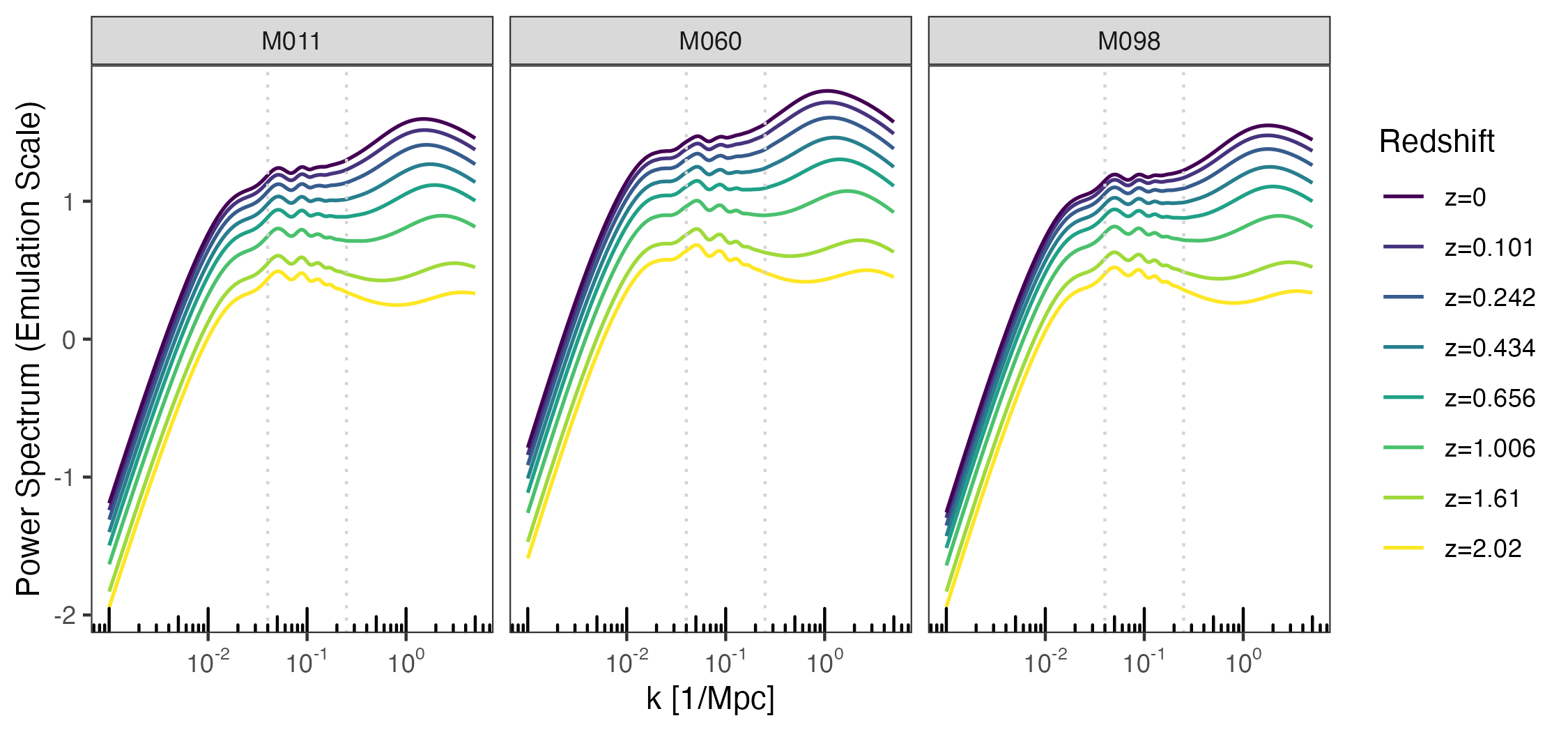}
	\caption{Learned mean MPS for three example cosmologies at each of the 8 redshifts.}
	\label{fig:dpc_cosmos}
	\end{center}
\end{figure}

The DPC model successfully learned the smooth matter power spectra from the suite of Mira-Titan simulations. After burn-in, the MCMC chain produced stable estimates for all spectra-time combinations. 
The smooth matter power spectra for one example cosmology is shown in Figure \ref{fig:data_mtnoisy_wmean}.
The learned spectra at multiple redshifts are shown in Figure \ref{fig:dpc_cosmos}.
The model effectively captures the BAO region's wiggles, which are of particular cosmological significance because they encode information about the universe’s expansion and dark energy effects.

Quantile-quantile (q-q) and residual plots are shown in Figures \ref{fig:dpc_cosmoqq} and \ref{fig:dpc_cosmosresid}, respectively.
MCMC diagnostics are provided in the Supplemental Materials; all were satisfactory. 

The Supplemental Materials also includes an assessment of the performance of the other methods considered in simulation relative to the DPC model, quantitatively comparing uncertainty interval widths and visually assessing their ability to model the shape of the power spectra; overall, the DPC model provides the best balance of accuracy and uncertainty management, particularly in regions of cosmological importance like the BAO.
Analagous q-q and residual plots for GAMs, heteroskedastic GPs, and deep GPs are also shown in the Supplemental Materials.
These models all have worse fits than the DPC model according to these diagnostics. 

\begin{figure}
	\begin{center}
    \includegraphics[width=0.9\textwidth]{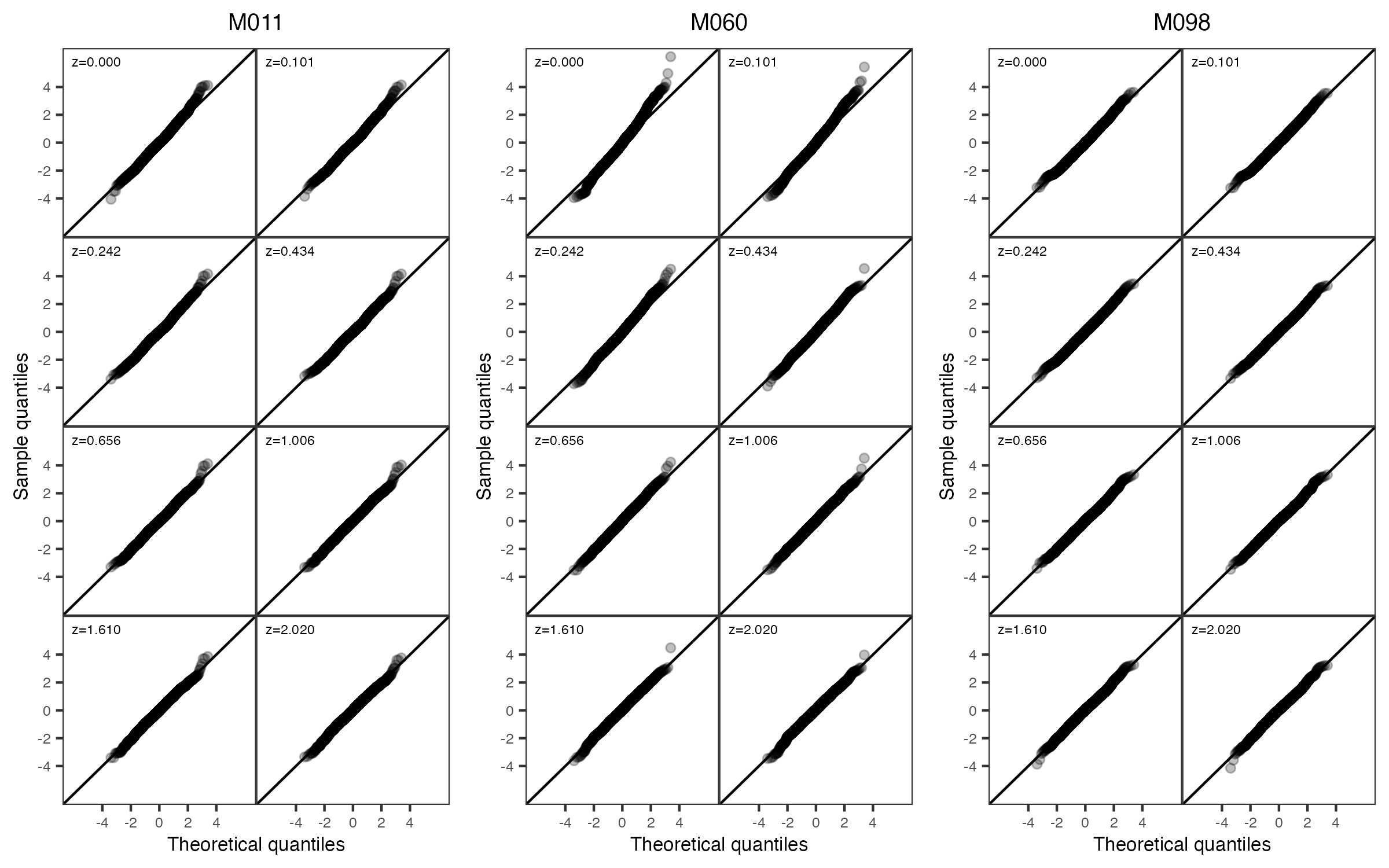}
	\caption{Quantile-quantile plots of the standardized residuals are shown for the lower- and higher-resolution simulations across eight redshifts for three cosmologies. These standardized residuals are calculated by subtracting the estimated mean from the simulations and then scaling each value by the square root of the resolution-dependent precision at its $k$ value. We expect the resulting sample to follow a standard normal distribution independent of $k$. The q-q plots compare the sample quantiles of these standardized residuals to the theoretical quantiles of a standard normal distribution. While the tails of the distributions show some divergence from the $x=y$ line, indicating slightly heavier tails than expected under normality, the overall model fit is relatively good.}
	\label{fig:dpc_cosmoqq}
	\end{center}
\end{figure}

\begin{figure}
	\begin{center}
    \includegraphics[width=0.9\textwidth]{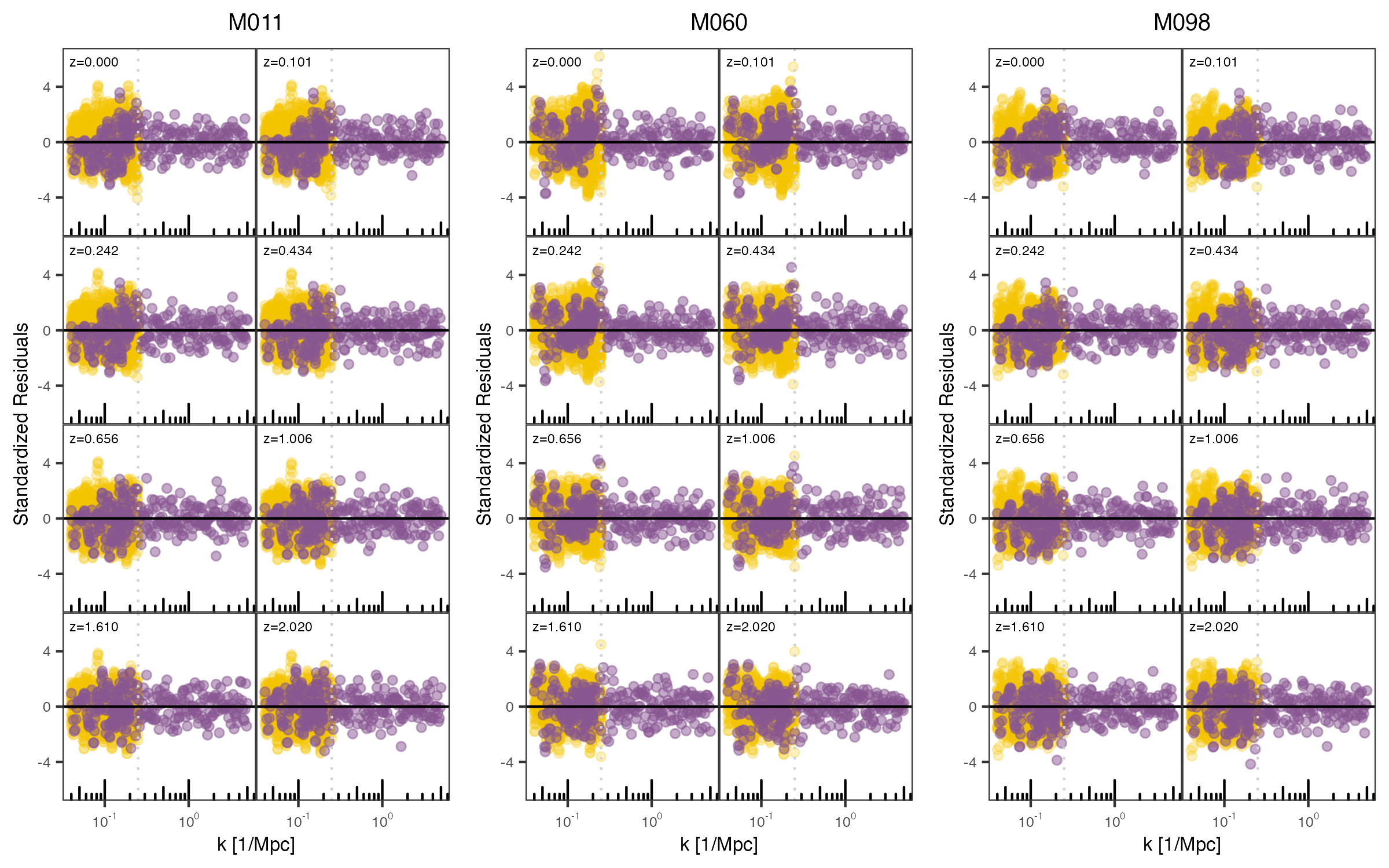}
	\caption{Standardized residuals for both the LR (yellow points) and HR (purple points) simulations at the eight redshifts for three cosmologies plotted against $k$. For certain cosmology and redshift combinations there is a slight trend in the HR residuals as we move from low $k$ to high $k$, up to the point where the LR simulations are excluded from the smoothing (dashed vertical line), indicating some potential bias between the LR and HR runs, likely due to box size effects in the N-body simulation. Aside from this trend, no obvious correlations or structures are visible.}
	\label{fig:dpc_cosmosresid}
	\end{center}
\end{figure}





\subsubsection{Learned Bandwidth}

\begin{figure}
	\begin{center}
    \includegraphics[width=0.6\textwidth]{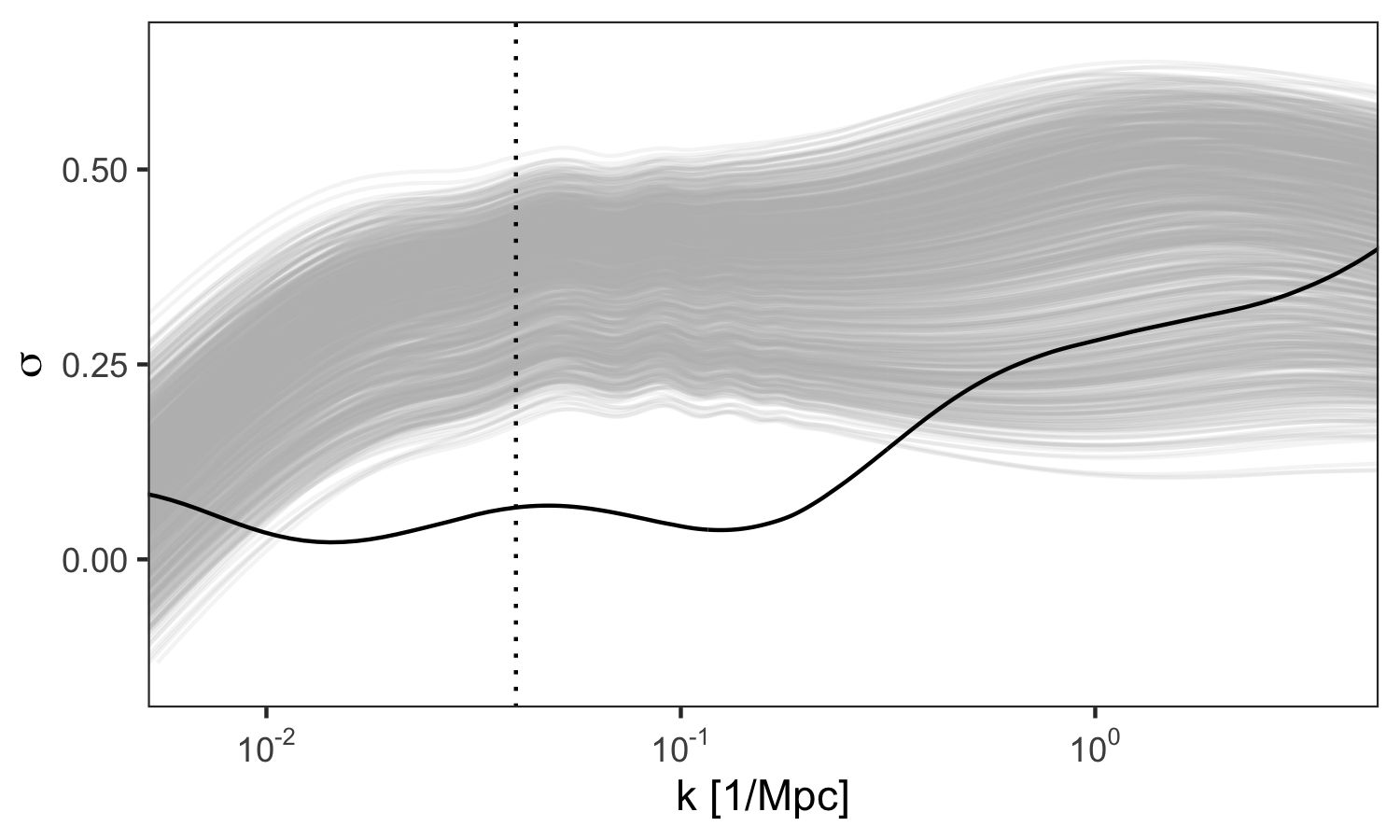}
	\caption{The median MCMC draw of the bandwidth function $\sigma$ (black line). Smaller $\sigma$ values indicate regions of the domain where the spectra are less smooth; $\sigma$ is small in the BAO region and large at high $k$ values, as anticipated. Also shown are the power spectra from all 111 cosmologies at each of the 8 redshifts, depicted in light gray and scaled to $\sigma$'s range. A vertical line indicates where PT meets simulation outputs.}
	\label{fig:sigma}
	\end{center}
\end{figure}

One notable feature of the results is the model’s ability to differentiate between the linear and nonlinear regions of the power spectrum. As shown in Figure~\ref{fig:sigma}, the bandwidth parameter $\sigma$ varies significantly across the domain. Smaller values of $\sigma$ correspond to less smooth regions, such as the BAO, where rapid oscillations in the power spectrum are expected. In contrast, the regions outside the BAO, especially at low k-values, show much smoother behavior, with larger $\sigma$ values indicating less variation.

This variation in $\sigma$ across different regions allows the DPC model to adapt flexibly to both high- and low-frequency patterns in the data, which is essential in cosmological modeling. The ability to fit both linear and nonlinear structures using a single model marks a significant advancement in emulation techniques for large-scale simulations like Mira-Titan.

\section{Discussion}
One challenge in using process convolutions is the choice of the number and location of grid points.  For complex functions, more grid points are needed to capture the underlying function.  Increasing the number of grid points also increases the computational complexity.  Unfortunately, determining the minimum number of grid points to adequately capture the function may require several attempts, using standard model diagnostics tools (e.g. on the residuals) to determine if additional grid points may be required. 
An example of performing these diagnostic checks is included in the Supplemental Materials. 

Although rare, the authors note that if grid points are more dense than the data, the posterior mean and credible intervals may contain spurious features in areas with little data.  These features disappear when a different grid is used (generally using fewer grid points).  Unfortunately, through various simulations, the authors have been unable to ascertain a consistent cause of these spurious features.  Therefore, as with any model, researchers must perform appropriate model diagnostics to determine if the model is behaving appropriately, and see if any unusual features may be artifacts of the grid itself.  The authors emphasize that these cases have been rare in their experience, but advise the reader to be aware of these possibilities in implementation.

For applications in which there is a potential bias between the simulations at different resolutions, as we observed for the LR and HR simulations from the N-body output at some cosmology and red shift combinations, one could model the vertical offset as a constant shift at the join points where different simulation resolutions overlap.
In a Bayesian framework, call this vertical offset at the join points between LR and HR simulations \( \Delta \); one could place a prior on \( \Delta \), such as:
\[
\Delta \sim \mathcal{N}(0, \sigma_\Delta^2).
\]
The above prior could be used to reflect any assumptions about the magnitude and direction of the offset (e.g., small and centered around zero). The likelihood function would combine the LR and HR data, accounting for the shift at the join point:
\[
P(D \mid f_{\text{LR}}, f_{\text{HR}}, \Delta) \propto \prod_{k < k_{\text{join}}} \mathcal{N}(P_{\text{LR}}(k) \mid f_{\text{LR}}(k)) \prod_{k \geq k_{\text{join}}} \mathcal{N}(P_{\text{HR}}(k) \mid f_{\text{HR}}(k) + \Delta).
\]
Future improvements to the DPC model will allow users to infer a unique \( \Delta \) value at each join location as part of the posterior distribution. 
Note that another potential source of bias, \( k \)-dependent bias, would require a more complex model where the offset \( \Delta(k) \) varies with \( k \). One could, e.g., place a Gaussian process prior on a discrepancy term \( \Delta(k) \) to capture the dependence of the offset on the scale \( k \). 

The data were analyzed using an R package entitled `dpc' (deep process convolutions), that is available for download via \texttt{github.com/LANL}. 
The package implements the MCMC in the C programming language, and parallelizes the computation of the likelihood, if such parallelization is available on the machine.

\section*{Acknowledgements}

This work was in part supported by the U.S. Department of Energy, Office of Science, Office of Advanced Scientific Computing Research, Scientific Discovery through Advanced Computing (SciDAC) program. 
Approved for public release: LA-UR-24-25003

\bibliographystyle{apalike}
\bibliography{references}

\end{document}